G

\documentstyle[prl,aps,epsf,amssymb]{revtex}

\input  epsf

\begin{document}

\title{Non-ergodic effects in the Coulomb glass: specific heat}

\author{A.\ D\'\i az-S\'anchez,$^{1,2}$ A.\ M\"obius,$^1$ 
   M.\ Ortu\~no,$^2$ A.~Neklioudov,$^1$ and M.\ Schreiber$^3$}
\address{
$^1$Institut f\"ur Festk\"orper- und Werkstofforschung, 
D-01171 Dresden, Germany,\\
$^2$Departamento de F\'\i sica, Universidad de Murcia,
E-30071 Murcia, Spain,\\
$^3$Institut f\"ur Physik, Technische Universit\"at, 
D-09107 Chemnitz, Germany
}

\date{\today}
\maketitle

\begin{abstract}

We present a numerical method for the investigation of non-ergodic
effects in the Coulomb glass. For that, an almost complete set of
low-energy many-particle states is obtained by a new algorithm.
The dynamics of the sample is mapped to the graph formed by the
relevant transitions between these states, that means by transitions
with rates larger than the inverse of the duration of the measurement.
The formation of isolated clusters in the graph indicates
non-ergodicity. We analyze the connectivity of this graph in
dependence on temperature, duration of measurement, degree of
disorder, and dimensionality, studying how non-ergodicity is reflected 
in the specific heat.

\end{abstract}


\section{Introduction}

Disordered systems of interacting localized particles have been
extensively studied for over two decades. A characteristic feature of
these systems is a complex many valley structure of the energy
landscape of the state space \cite{MP87}. Therefore, at sufficiently
low temperatures, the system cannot be considered to be in
thermodynamic equilibrium: Gibbs ensemble theory of statistical
mechanics, which is based on the equivalence of time and ensemble
averages, is not applicable. Thus non-ergodic effects are important.

The Coulomb glass \cite{PO85,SE84} is a prominent example of such
disordered systems. In heavily doped crystalline semiconductors,
amorphous semiconductor-metal alloys, and granular metals,
it plays an important role as a semiclassical model for systems of
localized states. The dynamical behavior of the Coulomb glass has been
studied by several groups: Schreiber et al.\
\cite{ST94a,ST94b,ST94c,ST96}, as well as
P\'erez-Garrido et al.\ \cite{PO98,PO99} determined numerically the 
transition probabilities between low-energy many-particle states, and
studied the eigenvalues of the transition probability matrix. The
former group directly diagonalized this matrix, whereas the latter
developed a renormalization method to eliminate the transitions with
large rates, what considerably simplifies the diagonalization. A broad
distribution of relaxation times over several orders of magnitude was
found in both cases. It reflects the glassy behavior of this system.
Moreover, Wappler et al.\ \cite{WS97} used the damage-spreading
algorithm to study the temporal evolution of the system, and found
evidence for a dynamical phase transition. Yu studied the time 
development of the Coulomb gap considering a self-consistent equation 
for the density of states \cite{Y99}. She too observed that very long 
time scales are involved.

The main aim of this work is to study numerically non-ergodic effects
in the Coulomb glass analyzing the transitions between many-particle
states. We apply this procedure to the investigation of such effects
in the specific heat. The paper is organized as follows: Section II
introduces the Coulomb-glass model. Section III describes how the
low-energy many-particle states are obtained numerically. In Sec.\
IV, we calculate the transition probabilities between these states,
and map the dynamical behavior of the Coulomb-glass sample to a graph. 
The nodes of the graph represent the many-particle states, and the 
edges the relevant transitions between them. Analyzing the structure 
of this graph, we determine the value of a physical observable in 
dependence on the duration of its measurement. In Sec.\ V, 
we use this method for the investigation of the non-ergodic effects in 
the specific heat: We study the influence of temperature, duration of 
measurement, disorder, and dimensionality. Finally, in Sec.\ VI we
extract some conclusions.

\section{Models}

The classical impurity band (CIB) model is the most realistic model
for simulating an impurity band of localized states in a lightly doped
semiconductor when quantum interference can be neglected 
\cite{PO85,SE84,ES85}. It is applicable if the following two conditions 
are fulfilled: (i) The mean nearest-neighbor distance is considerably 
larger than the localization radius of the wavefunction of an isolated 
impurity state. (ii) The temperature is so low that both the activation 
to the conduction band / from the valence band, and the formation of 
doubly charged donors / acceptors can be neglected.

We consider a $d$-dimensional sample of an n-type, partially compensated 
semiconductor with donor concentration $N_{\rm D}$, and acceptor 
concentration $N_{\rm A}=K N_{\rm D}$. The degree of compensation $K$ 
can range from 0 to 1. So donors are either occupied, that means 
neutral, or empty, that means positively charged. Acceptors captured an 
electron each, and are negatively charged. The distribution of electrons 
between the donors is governed by the Hamiltonian
\begin{equation}
H = -\sum_{i\nu}\frac{1-n_i}{r_{i\nu}}+
\sum_{i<j}{\frac{(1-n_i)(1-n_j)}{r_{ij}}}\,.
\end{equation}
The donor occupation number $n_i$ equals 1 for occupied donors, and 0 
for ionized donors. Moreover, $r_{i\nu}= | {\bf r}_i-{\bf r}_\nu |$ and 
$r_{ij}= | {\bf r}_i-{\bf r}_j |$, where the random positions of
the donors are denoted by ${\bf r}_i$ and ${\bf r}_j$, and those of 
the acceptors by ${\bf r}_\nu$. However, to minimize size effects, we 
impose periodic boundary conditions and use the minimum image convention 
\cite{MR53}. That means, in computing $| {\bf r}_i-{\bf r}_j |$, we 
substitute the projection of $| {\bf r}_i-{\bf r}_j |$ onto each of the 
coordinate axis, $x_i^{(\delta)}-x_j^{(\delta)}$ with 
$\delta = 1,...,d$, by the smallest related value in a periodically 
repeated representation, $\mbox{Min} (|x_i^{(\delta)}-x_j^{(\delta)}|,
L -|x_i^{(\delta)}-x_j^{(\delta)}|)$ with $L$ being the edge size of the 
sample.  For numerical reasons, we construct the samples so that the 
nearest neighbor distance exceeds 0.5, a well justified approximation 
for amorphous semiconductors, but not for crystalline systems. In this 
work, the unit of distance is defined by the donor density 
$\rho \equiv 1$, and electron charge, dielectric constant, and 
Boltzmann constant are taken to be unity.

Within the present study, most of the calculations are performed for a
simplified version of this model. Following Refs.\
\onlinecite{BE79,DL84}, see also \onlinecite{PO85,SE84}, we consider
a partially filled band of elementary charges (particles) localized on a
regular lattice formed by the $N_{\rm D}$ donor 
sites. Here, the acceptors are substituted by background charges $-K$ 
at each of the lattice sites, guaranteeing electro-neutrality on average. 
The disorder is simulated by a random potential $\epsilon_i$. Its values 
are uniformly distributed between $-B/2$ and $B/2$. Thus, the 
influence of the randomness of the donor positions is ignored, as well 
as the correlations between the values of the acceptor potential at 
neighboring donor sites. Moreover, the rectangular distribution of the 
$\epsilon_i$ is a simplification neglecting contributions from 
particular close pairs of donors and acceptors. This model is 
represented by the Hamiltonian
\begin{equation}
H=\sum_i\epsilon_i n_i +\sum_{i<j} {\frac{(n_i-K)(n_j-K)}
{r_{ij}}}\,,
\end{equation}
where $n_i \in \{0,1\}$ denotes again the occupation number of site 
$i$. As above, $r_{ij}$ is the distance between sites $i$ and $j$ 
according to periodic boundary conditions. The lattice spacing is 
taken as unit of distance.

A mixed form of both the models (1) and (2) is obtained in the
following way: The sites are positioned at random, and the acceptor
potential is substituted by a random on-site potential plus
the potential of neutralizing charges $-K$ 
at each site \cite{DM98,PO97}.
This mixed form is more realistic than Eq.\ (2): it keeps the donor
disorder as the Hamiltonian (1), but simplifies the disorder 
contribution from the acceptors. By means of $\epsilon_i$, the 
influence of the random surroundings of host atoms in an amorphous 
semiconductor can be simulated. This model is in the following 
referred to as random-position-with-random-potential model.

The relaxation procedures, which we use, alternatively simulate the
sample to be isolated, or to be in contact with a particle
reservoir \cite{MR87}. The latter means that, instead of $H$, the
grand canonical potential, $h=H-\mu\sum_i n_i$, is minimized. The
value of the chemical potential $\mu$ depends on $K$, i.e., it is
fixed by the electro-neutrality condition. Here, we first obtain $\mu$
performing a canonical simulation with reduced accuracy. Then we
calculate the set of low-energy states, as described below, taking
into account particle exchange with the reservoir.

\section{Determination of a set of low-energy states}

For studying low-temperature properties treating correlations exactly, 
we need to know a set ${\cal S}$ of almost all many-particles states 
in a certain energy interval above the ground state energy. If the
occupation numbers are known, the energy of a state can easily and 
directly be determined from Eq.\ (1) because we are treating a classical 
system. The problems, however, consist in the binomially large number of 
possible configurations, and in the existence of many local minima. Thus 
it is a complicated task to obtain such a set of low-energy 
many-particle states. It has been approached by several methods: 
Mochena and Pollak \cite{MP91} developed an approximative 
renormalization-like procedure. Schreiber and Tenelsen 
\cite{ST93,Tene.Schr} used the Metropolis algorithm to collect 
low-lying states which seems to be favorable in comparison to the 
previous method \cite{TE93}. M\"obius and Pollak \cite{MP96}, and 
P\'erez-Garrido et al.\ \cite{PO97} used two-stage algorithms. In the 
first stage, they obtained sets of local minima by means of 
relaxation. For that, a complete search considering rearrangements of 
the site occupations including up to four sites, and an incomplete 
search concerning more complex rearrangements, built up of several 
low-energy one-electron hops (shifts), were performed in Refs.\ 
\onlinecite{MP96} and \onlinecite{PO97}, respectively. In the second 
stage, both groups completed the table of low-energy states by 
systematically investigating the neighborhood (in the configuration 
space) of each of these states. This neighborhood is defined via the 
accessibility within only one of the considered rearrangements.

Here, we find the set ${\cal S}$ of $N$ low-energy many-particle 
states by means of a three-stage algorithm; for a short preliminary 
description see Ref.\ \onlinecite{DM98}. In the first two stages, we 
create, and improve a ``backbone'' of ${\cal S}$, formed by metastable 
states, the number of which can be considerably smaller than $N$. 
Then, in the third stage, we complete ${\cal S}$ by systematically 
investigating the neighborhood of the states found. Our procedure, 
which includes sophisticated local search \cite{MR92}, and thermal 
cycling \cite{MN97}, is explained in detail in the following.

In the first stage, creating the backbone of ${\cal S}$, we repeatedly 
start from states chosen at random, and simulate quenching the sample 
(i.e., a rapid relaxation) by means of a local search procedure: In an 
iterative process, we search the neighborhood of the actual state for 
states of lower energy, and accept always the first such state found. 
The process stops when no lower neighboring state exists. 

Our local search algorithm \cite{MR92} ensures stability with respect 
to rearrangements concerning one up to four sites. Making use of the 
branch-and-bound idea in order to avoid unnecessary attempts to a 
large extent, it considers the following rearrangements: 
\begin{itemize}
\item[(a)] transition of one electron between the sample and a 
reservoir, 
\item[(b)] arbitrary one-electron hops within the sample, 
\item[(c)] rearrangements by performing simultaneously an (a) and a 
(b) transition, and 
\item[(d)] arbitrary two-electron hops within the sample. 
\end{itemize}
To ensure high efficiency of the simulations, the searches in (c) and 
(d) have to be restricted to a certain number of neighbors. For one-, 
two-, and three-dimensional systems, we consider the first 4, 8, and 26 
neighbors, respectively. (Though the number of possible rearrangements 
increases rapidly with the number of neighbors considered, the portion 
of long-range hops among the energy decreasing rearrangements is small 
due to the decreasing interaction strength.)

The second stage consists in extending, and improving this set of 
metastable states by thermal cycling \cite{MN97}. For that, a further
set ${\cal M}$ of metastable states is considered. Initially, it 
equals a subset of ${\cal S}$, containing the states of lowest energy. 
We cyclically choose one of the states from ${\cal M}$, and apply to it 
a Metropolis process with a certain temperature. This process, which 
is referred to as heating, is terminated, however, after a small 
number of successful steps. Thus a great part of the information on 
the ground state, gained within the 
previous cycles, is retained. Then we quench the sample by means of
the above local search algorithm. If the energy of the final state is 
lower than the energy of the initial state, the latter is substituted 
in ${\cal M}$. 

For each of the final states of these cycles, we check whether or not 
it is already contained in ${\cal S}$. If not, and if, moreover, the 
total number of states is smaller than its maximum value, we add 
it to ${\cal S}$. However, if the final state is not contained 
within ${\cal S}$, but ${\cal S}$ has been already ``filled'', we 
substitute the final state for the state in ${\cal S}$ of highest 
energy, provided the energy of the final state is lower than the energy 
of the latter.

Several details of thermal cycling should be mentioned: When 
starting, the temperature of the Metropolis process is chosen to be 
equal to the total energy gain in quenching a random state divided by 
the number of sites. In the course of the thermal cycling procedure, 
this temperature is reduced gradually. The cyclic procedure continues 
until its efficiency becomes so low that, within a ``reasonable'' 
number of cycles, the set ${\cal M}$ has not been improved further. In 
the heating, we consider only two kinds of rearrangements, namely (a) 
and (b). For $d = 1$, 2, and 3, the latter are restricted to the first 
4, 8, and 26 neighbors, respectively. During this process, we fix the 
occupation of those sites which have the same $n_i$ value in all states 
contained in $\cal M$ because these values obviously favor a low energy. 
Moreover, in order to diminish the portion of unsuccessful attempts in 
the Metropolis process, we further reduce the set of rearrangements 
considered: We tabulate the relevant rearrangements after each 
temperature change, as well as after adding a new state to ${\cal M}$.
As relevance criterion, we use the condition that, for at least one 
of the states in $\cal M$, the related energy change must be smaller 
than the temperature. In heating, we consider only the excitations 
stored in this table. Each heating is terminated after having performed 
50 rearrangements.

In the third stage, we complete the set ${\cal S}$ of low-energy 
states by systematically investigating the neighborhood (in the 
configuration space) of the states found \cite{PO97,MP96}: For each 
state in ${\cal S}$, we construct all neighboring states, considering 
the same types of rearrangements as above. However, we perform (c) and 
(d) searches (restricted also here to the first 4, 8, and 26 
neighboring sites if $d = 1$, 2, and 3, respectively) only for those 
states, which are local minima with respect to the (a) and (b) 
rearrangements. Each new state is added to ${\cal S}$ if the number 
of states in ${\cal S}$ is smaller than its maximum value. Otherwise, we 
substitute the highest-energy state in ${\cal S}$ provided the energy 
of the new state is lower. 

This completion procedure starts with the state which has the highest 
energy, and proceeds cyclically until all states contained in ${\cal S}$ 
have been investigated. If the number of states in ${\cal S}$ is 
large, the decision whether or not the current state has been added to 
${\cal S}$ already previously, is particularly CPU time consuming. 
This decision is significantly accelerated by searching hierarchically 
in an array ordered according to the energies, where pointer 
arithmetics is used.

To check our program, we performed a series of tests. The degree of 
completeness of the spectrum of many-particle states within a certain 
energy interval above the ground state energy was judged in a similar 
way as in Ref.\ \onlinecite{MP96}. We observed that the completeness 
and the CPU time required do not only depend on the number $N$ of 
states included in ${\cal S}$, and on the number of local searches 
performed in the first stage, but to a considerable extent also on the 
number of states in ${\cal M}$. In the numerical experiments described
in Sec.\ V, $N$ has values between $25\,000$ and $75\,000$ what ensures 
that the width of the related energy interval exceeds the temperature 
by at least a factor of 25. We found that it is a good choice to
adjust the number of local searches in the first stage to $N/25$, 
and the number of states in ${\cal M}$ to $N/500$. Naturally, the 
kinds of rearrangements considered in the three stages are very 
important too. 

We devoted special attention to testing the thermal cycling part, 
which improves the backbone of ${\cal S}$. Its efficiency in 
finding states of particularly low energy is illustrated by Fig.\ 1. 
This graph presents the mean energy of the lowest state found in 
dependence on the CPU time required, contrasting results for simulated 
annealing (same rearrangements as in heating), sophisticated 
multi-start local search (repeatedly quenching states chosen at random 
by means of rearrangements (a) - (d)), and thermal cycling. Fig.\ 1 
shows that, in searching for states of very low energy,
thermal cycling is not only far superior to simulated 
annealing, but also to the sophisticated multi-start local search 
algorithm. Thus the incorporation of thermal cycling leads to a 
considerable efficiency increase in the construction of sets of 
low-energy many-particle states.

As a check of our numerical approach to obtaining almost complete sets 
of low-energy many-particle states, we verified that the corresponding 
results for the equilibrium specific heat agree with Refs.\ 
\onlinecite{MP96,MT97}.

\section{Influence of the duration of measurement}

Our main aim is to study the influence of the duration $\tau_{\rm m}$ 
of the measurement, during which the state of the sample travels 
randomly through its configuration space, on the expectation value of an 
observable $O$. For that, we decompose the configuration space into 
separate regions, isolated from each other on the time scale 
$\tau_{\rm m}$, in other words, into clusters of states. Two 
approximations are basic for this approach: (i) The duration of 
measurement is long enough for establishing thermal equilibrium inside 
each cluster. (ii) Transitions between different clusters are so rare 
that they can be ignored on the time scale of the measurement. 

First, we calculate the transition probability per unit time from the 
many-particle state $I$ to the state $J$. According to Ref.\ 
\onlinecite{PO85}, it is given by
\begin {equation}
W_{I \rightarrow J}= w_0\, \exp \left(\frac{-2 \sum r_{ij}}{a} \right) 
\left\{ \begin {array}{c@{\ \ }c } 
\exp\left[(E_I-E_J)/T\right] & {\rm if}\ E_I<E_J \\
 1 & {\rm if}\ E_I>E_J\,
\end{array} \right. \,.
\end{equation}
In this equation, the parameter $w_0$ is a constant of the order of 
the phonon frequency, $w_0\sim 10^{13}\ {\rm s^{-1}}$. $a$ denotes the 
localization radius, and $E_I$ the energy of the state $I$. The sum 
term is an abbreviation of the minimized sum of the related hopping 
lengths; it concerns only those sites, the occupation of which is
changed in the transition. That means, we decompose the many-electron 
transition into independent one-electron hops. Among all possible such 
decompositions, we choose that for which the sum of the lengths of the 
one-electron hops takes its minimum value. 

Presuming the occupation probabilities of the states to have their
thermodynamic equilibrium values, we obtain the rate of transitions 
from state $I$ to state $J$,
\begin {equation}
R_{I \rightarrow J} = W_{I \rightarrow J}\, \exp \left(\frac{-E_I}{T} 
\right)\, Z^{-1} \,  
\end{equation}
where $Z$ is the partition function. In thermodynamic equilibrium,
$R_{I \rightarrow J} = R_{J \rightarrow I}$. Thus the transition time, 
i.e., the inverse of the equilibrium transition rate, is given by
\begin {equation}
\tau_{IJ} = \tau_0 \, \exp \left( \frac{2\sum r_{ij}}{a} + 
\frac{E_{IJ}}{T} \right)\,Z
\end{equation}
with $E_{IJ} = \max (E_I,E_J)$, and $\tau_0 = w_0^{-1}$.

Now, we map the dynamics of the Coulomb glass sample simulated to 
a graph, where the nodes represent the states, and the edges those 
transitions between them for which $\tau_{IJ} < \tau_{\rm m}$. 
The nodes which are directly or indirectly connected with each other by 
such edges form clusters. These cluster correspond to regions of the 
configuration space being isolated from each other on the time scale 
$\tau_{\rm m}$. Here we presume equilibrium in determining $\tau_{IJ}$. 
However, this principle can easily be generalized to the non-equilibrium 
case, see below.

Provided, at the beginning of the measuring process, the sample is 
in one of the states of the cluster $\alpha$, we measure as value of 
the observable $O$,
\begin{equation}
\langle O \rangle_\alpha(T) =  \sum_{I \in \alpha} O_I 
\exp \left(\frac{-E_I}{T} \right)\, Z_\alpha^{-1}\, ,
\end{equation}
where $O_I$ denotes the value of $O$ for the state $I$, and $Z_\alpha$ 
the partition function of this cluster. The values of 
$\langle O \rangle_\alpha$ can have a very broad distribution.

Finally, we turn to the expectation value of $O$, that is the mean value 
of repeated measurements, referred to in the following as 
$\langle\langle O \rangle\rangle(T,\tau_{\rm m})$. The probability to 
find the sample in one of the states belonging to the cluster $\alpha$ 
be $P_\alpha$. Thus, $\langle\langle O \rangle\rangle$ is given by the 
weighted average of the $\langle O_\alpha \rangle$:
\begin{equation}
\langle\langle O \rangle\rangle(T,\tau_{\rm m})= 
\sum_\alpha \langle O \rangle_\alpha \, P_\alpha \,.
\end{equation}
In consequence, $\langle\langle O \rangle\rangle$ depends via the cluster 
structure and $P_\alpha$ on $\tau_{\rm m}$, and also on $T$. 

Both the cluster structure and $P_\alpha$ are influenced by the 
history of the sample. In our numerical experiments, we have 
simulated two situations:
\begin{itemize}
\item[(A)] 
As standard case, we presume that the sample has reached thermal
equilibrium before the measurements; thus $P_\alpha = Z_\alpha /Z$.
\item[(B)]
Alternatively, we assume the sample to have been quenched from 
infinite $T$ to the measuring $T$ within a short time interval 
$\tau_{\rm q}$. To emulate this process we quench first to $T=0$, and 
heat then immediately to the measuring $T$: In the beginning, we 
assign the same probability to all states, $P_I = 1/N$. Then, we perform
the following iteration treating the states according to decreasing 
energy. We re-distribute the weight of the considered state
to all those states of lower energy to which a transition can happen
within $\tau_{\rm q}$ modifying the set of the $P_I$ correspondingly. 
The related transition times are obtained in a way analogous to the 
derivation of Eq.\ (5). That means, substituting the equilibrium 
occupation probabilities by the (iteration cycle dependent) 
non-equilibrium $P_I$, we get for the transition $I \rightarrow J$
\begin{equation}
\tau_{I \rightarrow J}=
\left\{ \begin {array}{c@{\ \ }c } 
\infty & {\rm if}\ E_I<E_J \\
w_0^{-1} \exp \left( 2\sum r_{ij}/a \right)\, P_I^{-1} & 
{\rm if}\ E_I>E_J\,
\end{array} \right. \,.
\end{equation}
Thus the quenching process is related to the modification of parts of 
the states graph (differing from the ``equilibrium graph'') in each of 
the iteration steps. At the end of the iteration, only the local minima 
have a finite occupation probability. Finally, we assign to each 
``equilibrium cluster'' the sum of the occupation probabilities of the 
included ``non-equilibrium local minima''. In this way, we calculate the
non-equilibrium values of $P_\alpha$, but we neglect
the influence of the preparation mode on $\tau_{IJ}$, and thus on the
structure of the clusters which are isolated from each other on the time
scale $\tau_{\rm m}$.  
\end{itemize}

\section{Non-ergodic specific heat}

Applying the methods presented in Secs.\ III and IV to the study of 
non-ergodic effects in the specific heat, we start from the definition 
of this observable for a cluster $\alpha$ of states, which is in 
thermal equilibrium:
\begin{equation}
c_\alpha(T) = 
\frac{\langle H^2 \rangle_\alpha - \langle H \rangle_\alpha^2}
{T^2 N_{\rm D}} \,.
\end{equation}
Since the specific heat is by itself a quantity relating to an ensemble 
of states rather than to a single state, we do not need brackets here to
mark the averaging over the states in the cluster $\alpha$, in deviation 
from the general notation $\langle O \rangle_\alpha(T)$ in Eq.\ (6). 
After averaging over the set of clusters, the $\tau_{\rm m}$ 
dependent value of the specific heat $\langle c \rangle(T,\tau_{\rm m})$ 
is obtained utilizing Eq.\ (7). 

To make the influence of $\tau_{\rm m}$ directly visible, we consider 
the quotient of the values of the specific heat for finite and infinite 
duration of measurement, respectively:
\begin{equation}
q(T,\tau_{\rm m}) = \frac
{\langle c \rangle(T,\tau_{\rm m})}{\langle c \rangle(T,\infty)}\,.
\end{equation}
This quantity is in the following referred to as non-ergodicity 
quotient. Studying $q(T,\tau_{\rm m})$ rather than 
$\langle c \rangle(T,\tau_{\rm m})$ is additionally
motivated by numerical reasons: The random fluctuations of $q$ from 
sample to sample are considerably smaller than the fluctuations of 
$\langle c \rangle$. Moreover, in order to characterize 
the fluctuations of $c$ from measurement to measurement, we consider its
root-mean-square deviation, related to the distribution of the clusters,
\begin{equation}
\sigma^2(T,\tau_{\rm m})= \sum_\alpha P_\alpha \, \left(
\frac 
{c_\alpha(T,\tau_{\rm m}) - \langle c \rangle(T,\tau_{\rm m})}
{{\langle c \rangle(T,\infty)}\,}
\right)^2 
\,.
\end{equation}

Investigating the physical properties of macroscopic systems, we have 
calculated ensemble averages, and have compared the results for different 
sample sizes. The number of samples to be taken into account depends not 
only on the accuracy to be achieved, but also on all the various details 
of the simulated situation. Generally, we have chosen the size of the 
samples so large that corresponding convergence of the results is ensured. 
The related parameter values are given in the figure captions.

Several details of our simulations have to be mentioned: The
localization radius $a$ equals always 0.2 \cite{MP91b}. Decreasing its 
value corresponds, crudely speaking, to stretching the time scale, and 
thus enhances the non-ergodic effects. Since a great part of the 
literature concerns the simplified Coulomb glass model according to Eq.\ 
(2) (sites arranged on a regular lattice), we, too, consider this 
situation in most of the simulations. Thus, if not stated otherwise, our 
numerical experiments have been performed for this model. The results 
presented here have been obtained by means of a canonical procedure: In 
calculating $\langle c \rangle$, we take into account only the 
low-energy states with total charge equal to $K N_{\rm D}$. When the 
grand canonical procedure is used for this aim, finite-size effects are 
more important since, due to differing total charge, some clusters 
remain separated from each other in configuration space even if 
$\tau_{\rm m} \rightarrow \infty$.

In the following, we discuss our numerical results in detail. First, we 
consider the non-ergodicity quotient $q$. As examples, 
Figs.\ 2a and 2b illustrate for three-dimensional samples with two 
different degrees of disorder, how finite-size effects influence the 
dependence of $q$ on the duration of measurement $\tau_{\rm m}$. They 
show that finite-size effects are particularly important for large 
$\tau_{\rm m}$, where extended clusters of states have to be considered. 
However, if the sample size $N_{\rm D}$ exceeds a certain value, roughly 
500 for the cases considered here, $q$ is almost independent of 
$N_{\rm D}$. Certainly, this limit depends on the concrete situation 
considered, i.e., on $T$, $B$, and $d$, and also on $\tau_{\rm m}$. 

Moreover, Fig.\ 2 leads to an important conclusion: The value of 
$\tau_{\rm m}$ where $q$ reaches (almost) 1 can be interpreted as 
relaxation time of the specific heat. It exceeds typical experimental 
times by several orders of magnitude. This indicates glassy behavior, 
where non-ergodic effects are fundamental. 

In Fig.\ 3, we show $q(\tau_{\rm m})$ and its fluctuation $\sigma$ 
defined by Eq.\ (11). These data illustrate that the measured values of 
$c$  are distributed within a wide range around the mean value, what 
originates from the varying properties of the separated clusters of 
states. This is a manifestation of the non-ergodic effects. However, for 
drawing further conclusions, a detailed study of the size dependence of 
$\sigma$ would be needed.

Figure 4 displays the temperature dependence of the specific heat for 
two values of $\tau_{\rm m}$, as well as for thermal equilibrium 
(infinite $\tau_{\rm m}$). Moreover, this figure compares the changes 
caused by variation of $T$ with the influence of $\tau_{\rm m}$.

Now we study the non-ergodicity quotient $q$ in more detail. Fig.\ 5 
illustrates that the strength $B$ of the disorder has only a weak 
influence within the considered parameter range. Only for $B = 1$, a 
small systematic shift of $q(\tau_{\rm m})$ towards higher 
$\tau_{\rm m}$ is detectable in comparison to the curves for stronger 
disorder. If $\tau_{\rm m}$ is so large that $q$ reaches almost 1, 
decreasing the disorder enhances the relevance of long-time 
correlations.

The influence of the temperature $T$ on $q$ is considered in Fig.\ 6. 
For $\tau_{\rm m}/\tau_0$ exceeding $10^{10}$, we observe the expected 
behavior, namely that decreasing $T$ is related to the increasing 
influence of non-ergodicity effects. However, we found a point, 
$\tau_{\rm m}/\tau_0 \sim 10^{10}$, where $q$ is almost independent of 
$T$ within the $T$ region studied. For smaller $\tau_{\rm m}$, there is 
even some increase of $q$ with decreasing $T$. Presumably, the reason of
this feature is the following. Decreasing $T$ influences the cluster
structure in two ways. On the one hand, it causes a division of clusters
due to the decrease of transition rates, and thus an increase of the
cluster number. On the other hand, however, the number of really 
relevant clusters decreases due to the decreasing occupation probability 
of excited states. We suppose that the second effect is dominating for 
small $\tau_{\rm m}$. There is some analogy to uncorrelated hopping 
conductivity, which increases exponentially with $T$ in the limit
$\omega \rightarrow 0$, but decreases as $1/T$ at high frequency 
\cite{P71}. A similar behavior has also been found in the frequency 
dependence of the dielectric susceptibility \cite{DO99}.

The question to which extent the obtained results are model dependent is 
answered in Fig.\ 7. For the chosen disorder strength, the results for 
the  CIB model and the random-position-with-random-potential model are 
almost identical. For the lattice model, however, the $q(\tau_{\rm m})$ 
curve is a bit steeper, but the relaxation time of the specific heat is 
roughly the same as for the other models. Thus, in the region of low 
$q$, its $q(\tau_{\rm m})$ curve is shifted towards higher 
$\tau_{\rm m}$. This effect probably originates from the missing of 
``easy'' hops between closely neighboring sites. Thus the influence of 
the spatial disorder of the donor sites is clearly stronger than the 
influence of the details of the random on-site potential $\epsilon_i$. 
(In the CIB model the acceptors create a potential which is almost 
Gaussian with a width of 2.5.)

We have also studied the non-ergodicity quotient $q$ for one- and 
two-dimensional samples. The results, presented in Fig.\ 8, are similar 
to the above findings for the three-dimensional case. However, there is 
a clear trend of non-ergodicity effects becoming more and more 
pronounced with decreasing dimension in the region of large 
$\tau_{\rm m}$. Thus the reduction of the dimension seems to be related 
to a stronger ``localization'' of the particles.

Finally, we consider the question how the sample preparation influences
$q(\tau_{\rm m})$. Fig.\ 9 shows results for the lattice model. Here, we 
compare samples being in thermal equilibrium with samples prepared by 
quenching. The latter curve is shifted towards larger $\tau_{\rm m}$. 
However, this effect decreases with increasing $B$, compare with Fig.\ 2 
in \cite{DM98}. Thus it is natural that the preparation conditions have 
a weaker influence for the other two Coulomb glass models considered.

\section{Conclusions}

In the previous sections, we have presented a numerical algorithm for 
studying the non-ergodic effects in the Coulomb glass, which are very 
important at low temperatures. This method considers many-particle 
states, and takes into account correlations completely. Here, it has 
been used for investigating how the mean value of the specific heat 
$\langle c \rangle$ depends on the measuring time. The main results of 
our numerical experiments are summarized in the following points:
\begin{list}{(\roman{enumi})}{\usecounter{enumi}}
\item The non-ergodicity quotient $q(T,\tau_{\rm m}) = 
\langle c(T,\tau_{\rm m}) \rangle / \langle c(T,\infty) \rangle$ 
strongly depends on the measuring time $\tau_{\rm m}$. It vanishes as 
$\tau_{\rm m} \rightarrow 0$, and, in the cases studied here, it 
approaches 1 only for values of $\tau_{\rm m}$ which exceed realistic
measuring times by orders of magnitude.
\item For large $\tau_{\rm m}$, $q$ decreases with decreasing $T$, 
whereas an increase of $q$ with decreasing $T$ is observed in the 
region of small $\tau_{\rm m}$.
\item Spatial disorder has a larger influence on the non-ergodic 
effects than the strength of the on-site random potential in the 
lattice model.
\item The importance of non-ergodic effects increases with decreasing
dimensionality.
\item Preparing the sample by a quench causes an increase of 
non-ergodic effects in the case of weak disorder.
\end{list}

Finally, we would like to stress one technical aspect of the
simulations. We have improved our previously developed algorithm for 
the construction of almost complete sets of low-energy many-particle 
states \cite{MP96} by adding a thermal cycling \cite{MN97} part. This 
method, originally applied to the traveling salesman problem, combines 
ideas from Monte Carlo and local search algorithms. Here, we have used
it when searching for deep local minima in the configuration space of 
the Coulomb glass. Also in this case, thermal cycling has been proved 
to be highly efficient.

\section*{Acknowledgments}

This work was supported by the SMWK and DFG (SFB 393). A great part of
it was performed during A.~D.-S.'s visit at the IFW Dresden; A.~D.-S.\ 
thanks the IFW for its hospitality. We are indebted to M.~Pollak for 
stimulating discussions.

\pagebreak

\begin{figure}
\caption{
Relation between CPU time, $\tau_{\rm CPU}$ (in seconds), and mean
deviation of the energy of the lowest state found from the ground state
energy, $\delta E = E_{\rm mean}-E_{\rm ground\ state}$, for one 
realization of the three-dimensional lattice model with $B = 1$ and
$N_{\rm D} = 1000$. $\times $: simulated annealing; $\triangle $: 
sophisticated multi-start local search; $\bullet $: thermal cycling. The 
values of $\tau_{\rm CPU}$ relate to one 180 MHz PA8000 processor of an 
HP K460. For simulated annealing and multi-start local search, averages 
were taken from 20 runs, for thermal cycling from 100 runs. In thermal 
cycling, the ground state was always found within 500 seconds. Error 
bars ($1\sigma$ region) are presented if they are larger than the 
symbols. The interpolating lines are guides to the eye only.
}
\label{fig1}
\end{figure}

\begin{figure}
\caption{
Influence of the sample size $N_{\rm D}$ on the dependence of the 
non-ergodicity quotient $q$ on the duration of the measurement $\tau_m$ 
for the three-dimensional lattice model. (a) $B=1$, $T=0.006$; (b) 
$B=4$, $T=0.008$. $\circ$, +, and $\bullet$ correspond to 
$N_{\rm D} = 216$, 512, and 1000, respectively. The ensemble averaging 
took into account 200 samples in each case, except for the two curves 
for $N_{\rm D} = 512$ and 1000 in (a),  where we considered 
100 and 50 samples, respectively. The error bars are smaller than the 
symbol size. 
}
\label{fig2}
\end{figure}

\begin{figure}
\caption{Fluctuation of the contributions to $q$ from the individual 
clusters, represented by error bars displaying $\sigma$ defined by Eq.\ 
(11). $B=1$, $T=0.008$, and $N_{\rm D} = 512$. Ensemble averaging took 
into account 100 samples. 
}
\label{fig3}
\end{figure}

\begin{figure}
\caption{Temperature dependence of the mean specific heat for thermal 
equilibrium ($\circ$), and for two finite measuring times: 
$\tau_{\rm m}/\tau_0=10^{15}$ ($\blacktriangle$) and 
$\tau_{\rm m}/\tau_0=10^{12}$ ($\times $). The error bars represent 
the width of the distribution of $c_\alpha$. $d = 3$, $B=1$ and 
$N_{\rm D} = 512$. Ensemble averaging took into account 100 samples. 
}
\label{fig4}
\end{figure}

\begin{figure}
\caption{
Influence of the disorder strength $B$ on $q(\tau_{\rm m})$ for the 
lattice model: $\bullet $: $B=1$; +: $B=2$; $\circ $: $B=4$. For all
curves, $d=3$, $N_{\rm D} = 512$, $T = 0.008$. Ensemble averaging 
took into account 200 samples for $B = 2$ and $B = 4$, and 100 samples
for $B = 1$.  
}
\label{fig5}
\end{figure}

\begin{figure}
\caption{
Influence of the temperature $T$ on $q(\tau_m)$ for the lattice model: 
$\bullet $: $T=0.006$;  $\blacktriangle $: $T=0.008$; +: $T=0.010$;
$\circ $: $T=0.012$. For all curves: $d=3$, $B=1$, and 
$N_{\rm D} = 512$; ensemble averaging took into account 100 samples.
}
\label{fig6}
\end{figure}

\begin{figure}
\caption{
Comparison of the $q(\tau_m)$ obtained for the three different versions
of the Coulomb glass model introduced in Sec.\ II: +, $\circ $, 
and $\bullet $ denote lattice, CIB, and 
random-position-with-random-potential models, respectively. For the
first and the last model, $B = 2$. In all cases, $d = 3$, $T=0.008$, and 
$N_{\rm D} = 512$; ensemble averaging took into account 200 samples. 
}
\label{fig7}
\end{figure}

\begin{figure}
\caption{
Comparison of the $q(\tau_m)$ obtained for lattice models of different 
dimensions: $\circ $: $d=1$, $N_{\rm D}=50$; +: $d=2$, $N_{\rm D}=484$; 
$\bullet $: $d=3$, $N_{\rm D}=512$. In all cases, $B=1$, and $T=0.01$. 
Ensemble averaging took into account 200 samples for $d=1$ as well as 
for $d=2$, and 100 samples for $d=3$.
}
\label{fig8}
\end{figure}

\begin{figure}
\caption{
Influence of sample preparation on $q(\tau_m)$ for the three-dimensional 
lattice model: $\circ $: sample in thermal equilibrium; $\bullet $: 
sample quenched within $\tau_q = 10^{13}\,\tau_0 \sim 1\ {\rm s}$. For 
both curves: $B=1$, $T=0.01$, and $N_{\rm D} = 512$; ensemble averaging 
took into account 100 samples. 
}
\label{fig9}
\end{figure}

\pagebreak
\epsfxsize=\hsize
\begin{center}
\leavevmode
\epsfbox{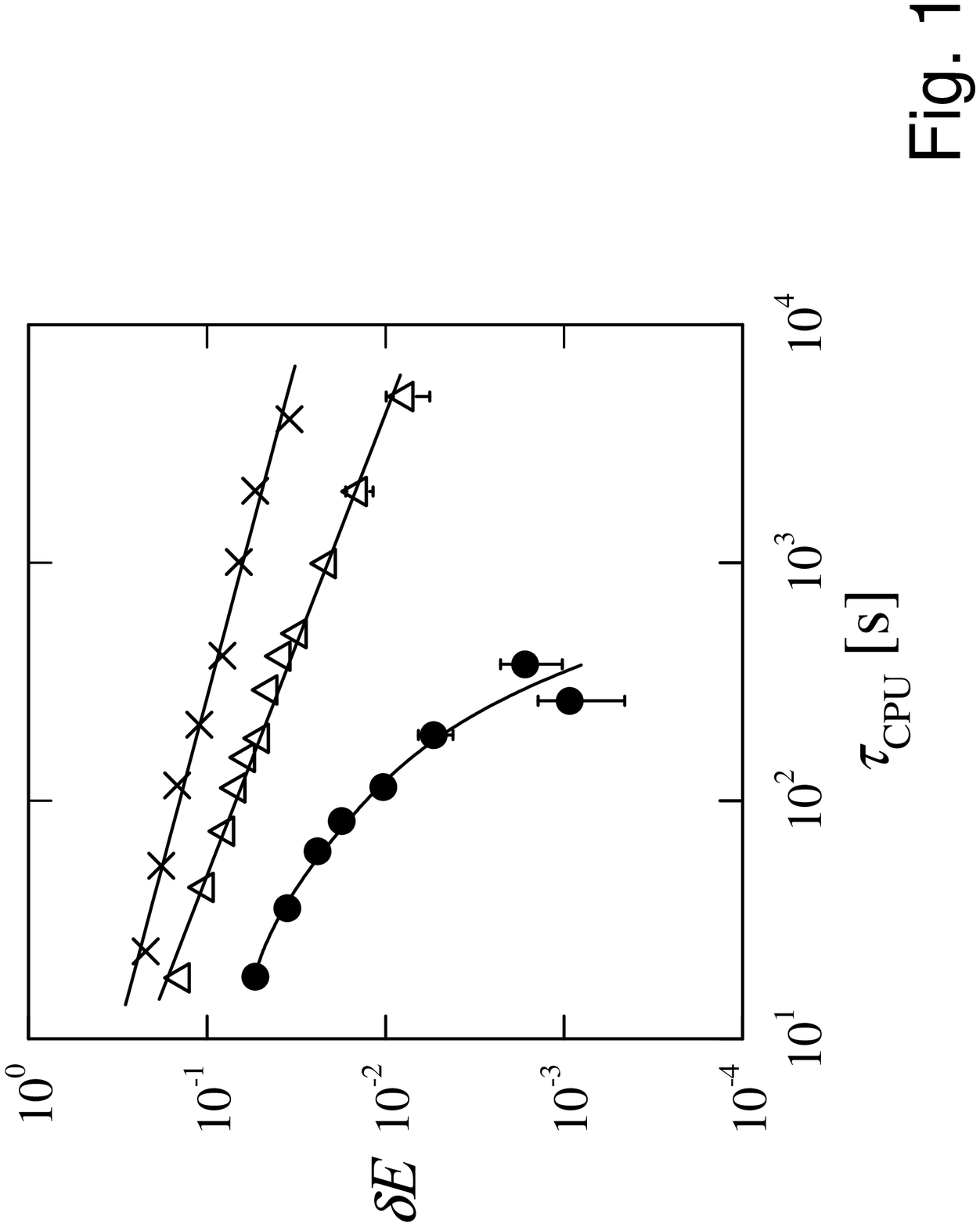}
\end{center}

\pagebreak
\epsfxsize=\hsize
\begin{center}
\leavevmode
\epsfbox{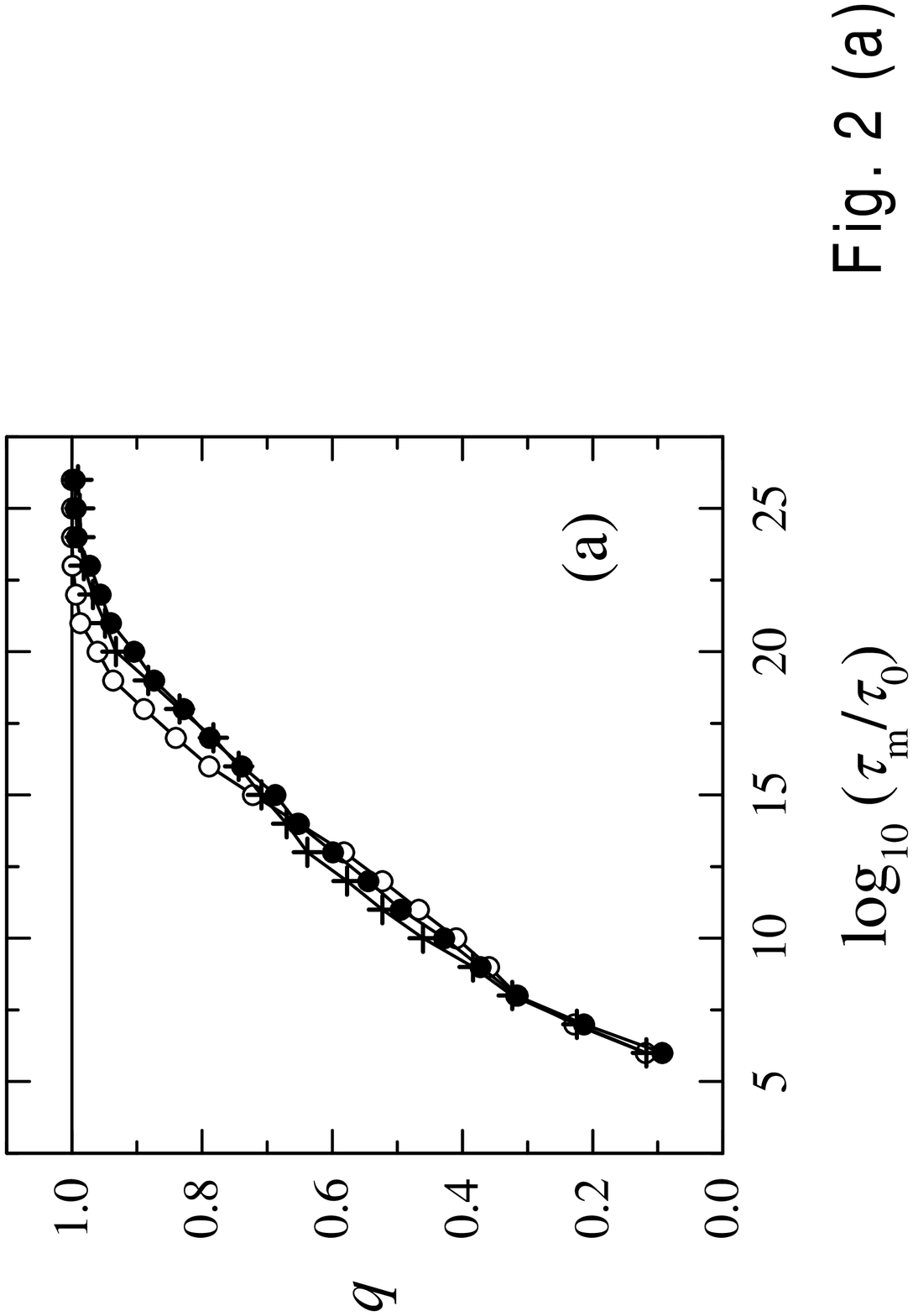}
\end{center}

\pagebreak
\epsfxsize=\hsize
\begin{center}
\leavevmode
\epsfbox{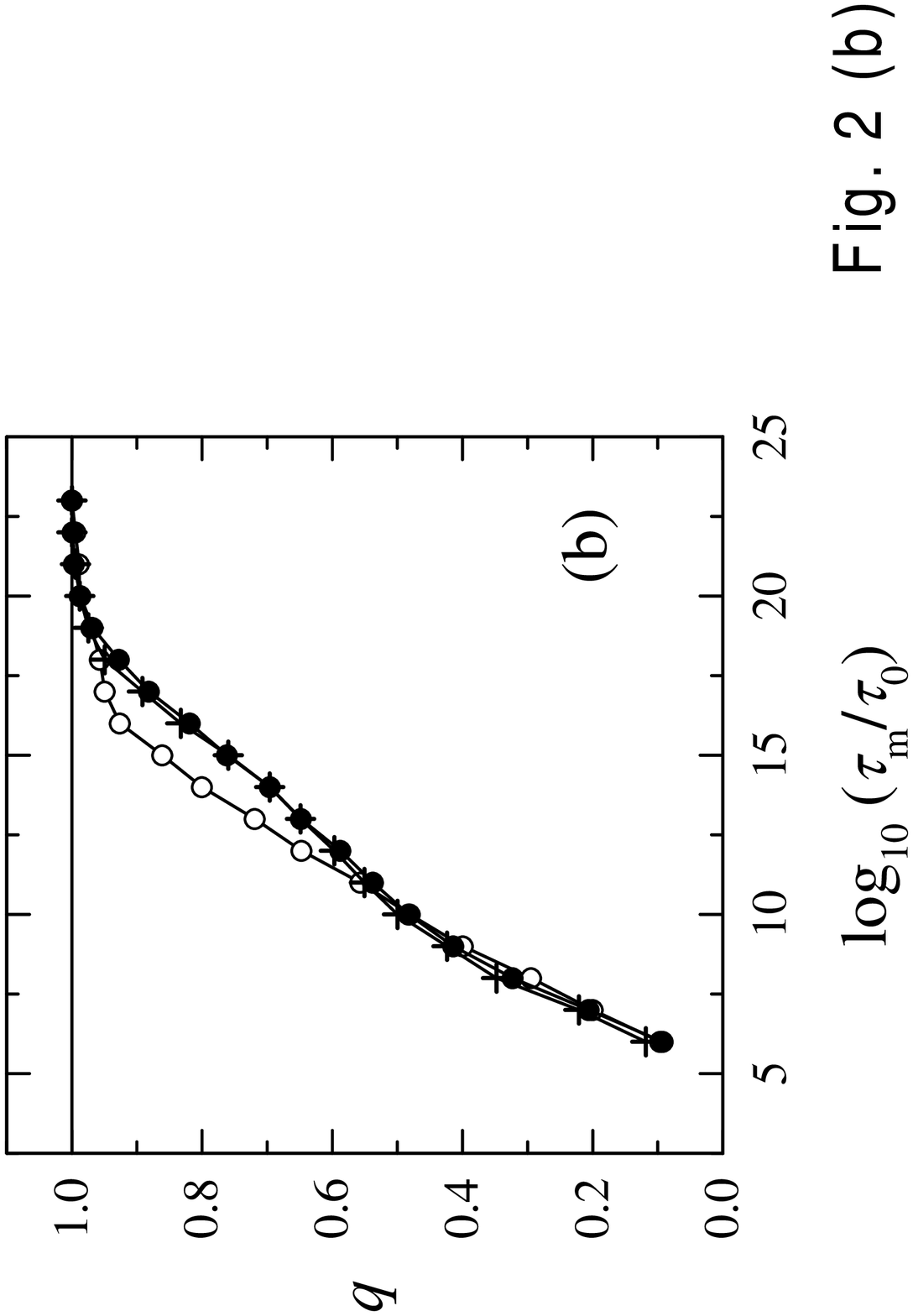}
\end{center}

\pagebreak
\epsfxsize=\hsize
\begin{center}
\leavevmode
\epsfbox{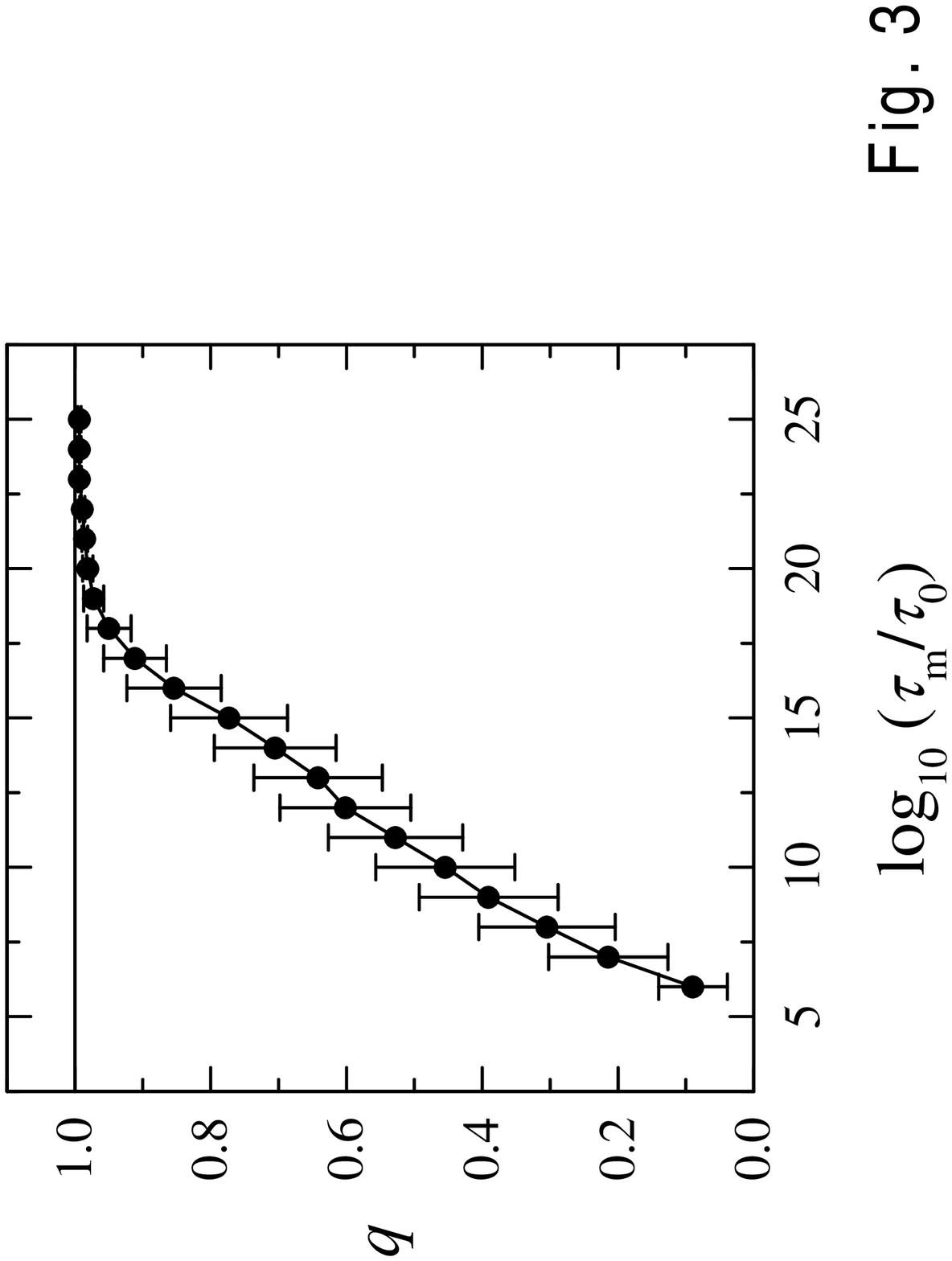}
\end{center}

\pagebreak
\epsfxsize=\hsize
\begin{center}
\leavevmode
\epsfbox{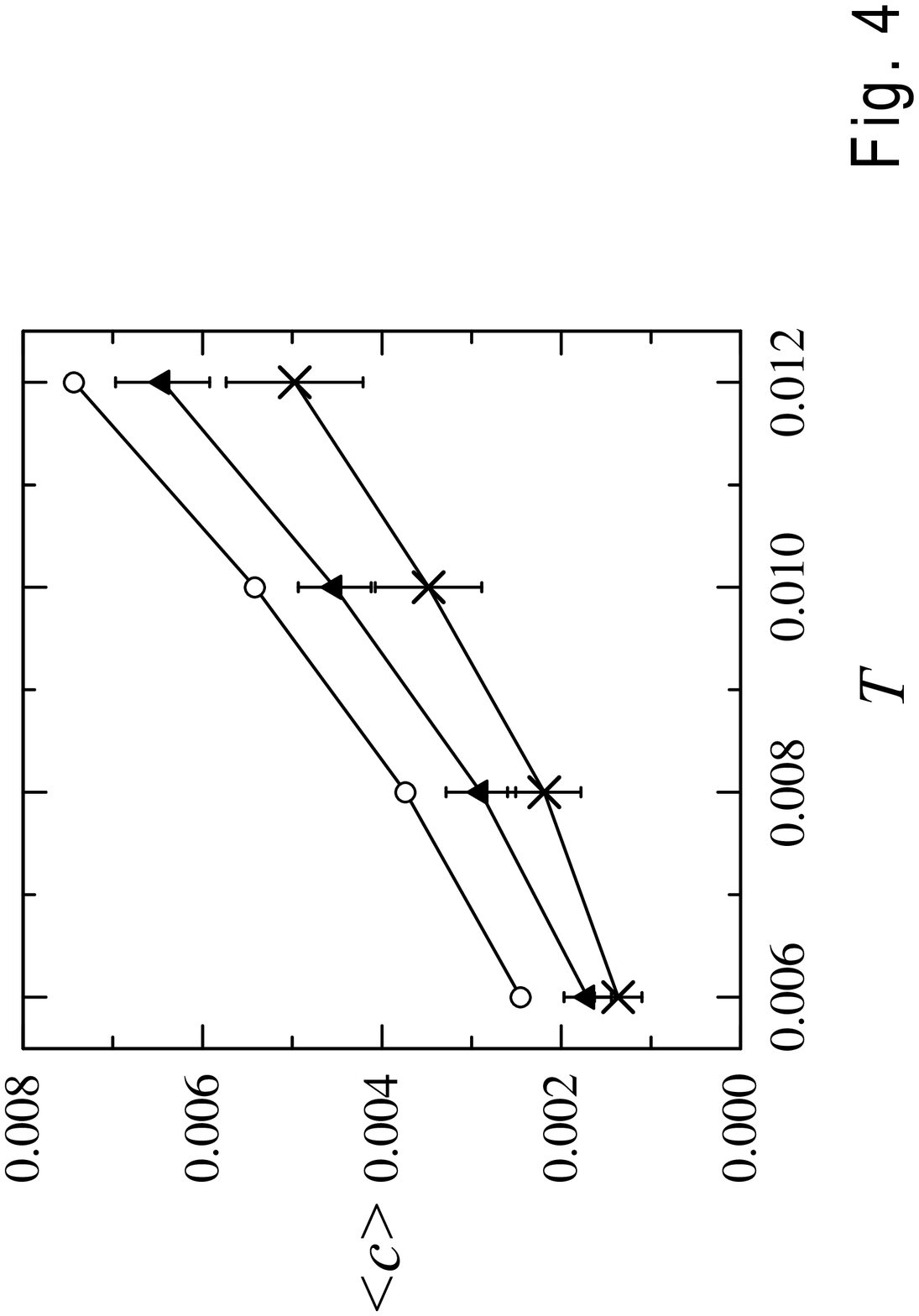}
\end{center}

\pagebreak
\epsfxsize=\hsize
\begin{center}
\leavevmode
\epsfbox{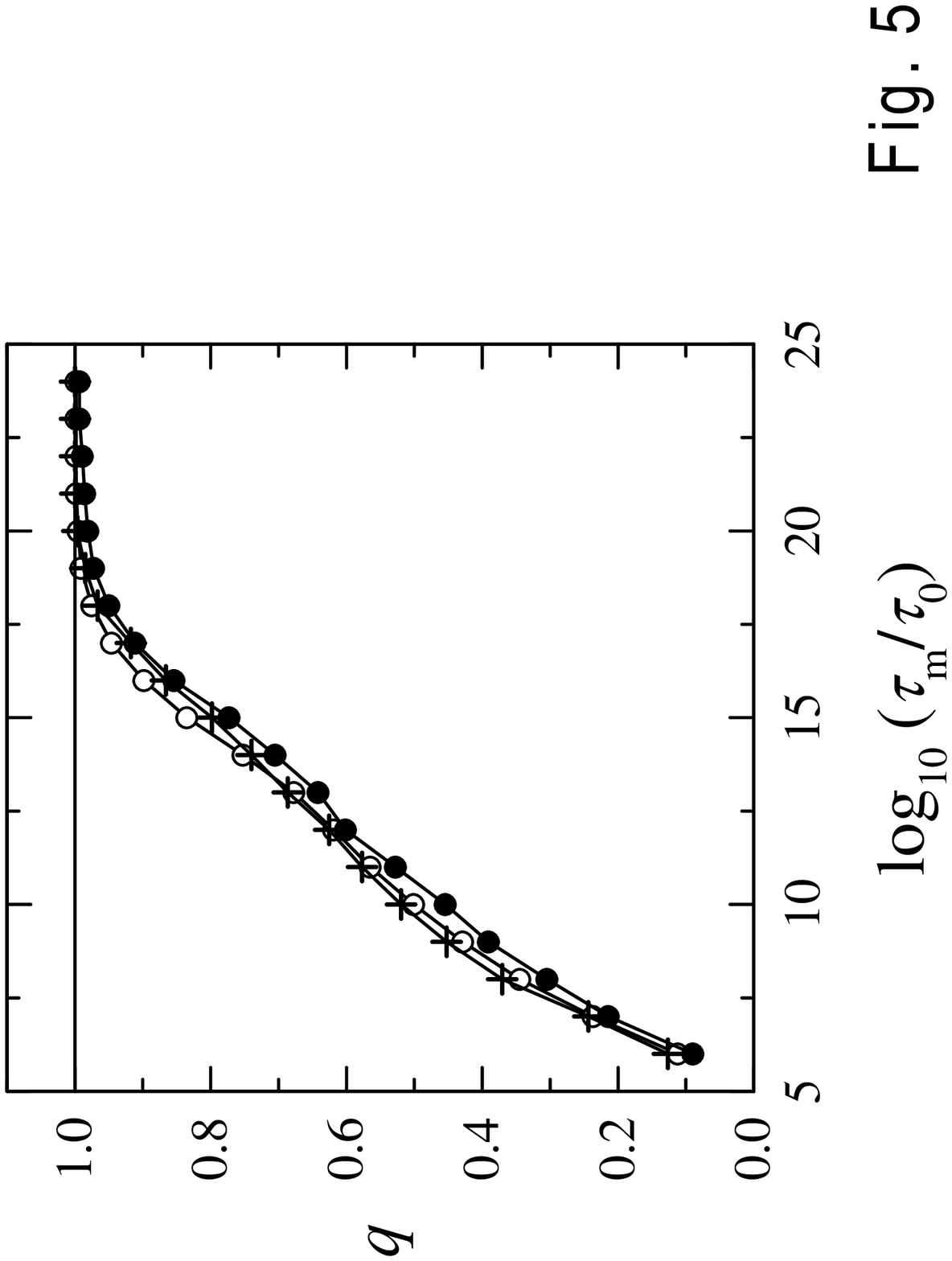}
\end{center}

\pagebreak
\epsfxsize=\hsize
\begin{center}
\leavevmode
\epsfbox{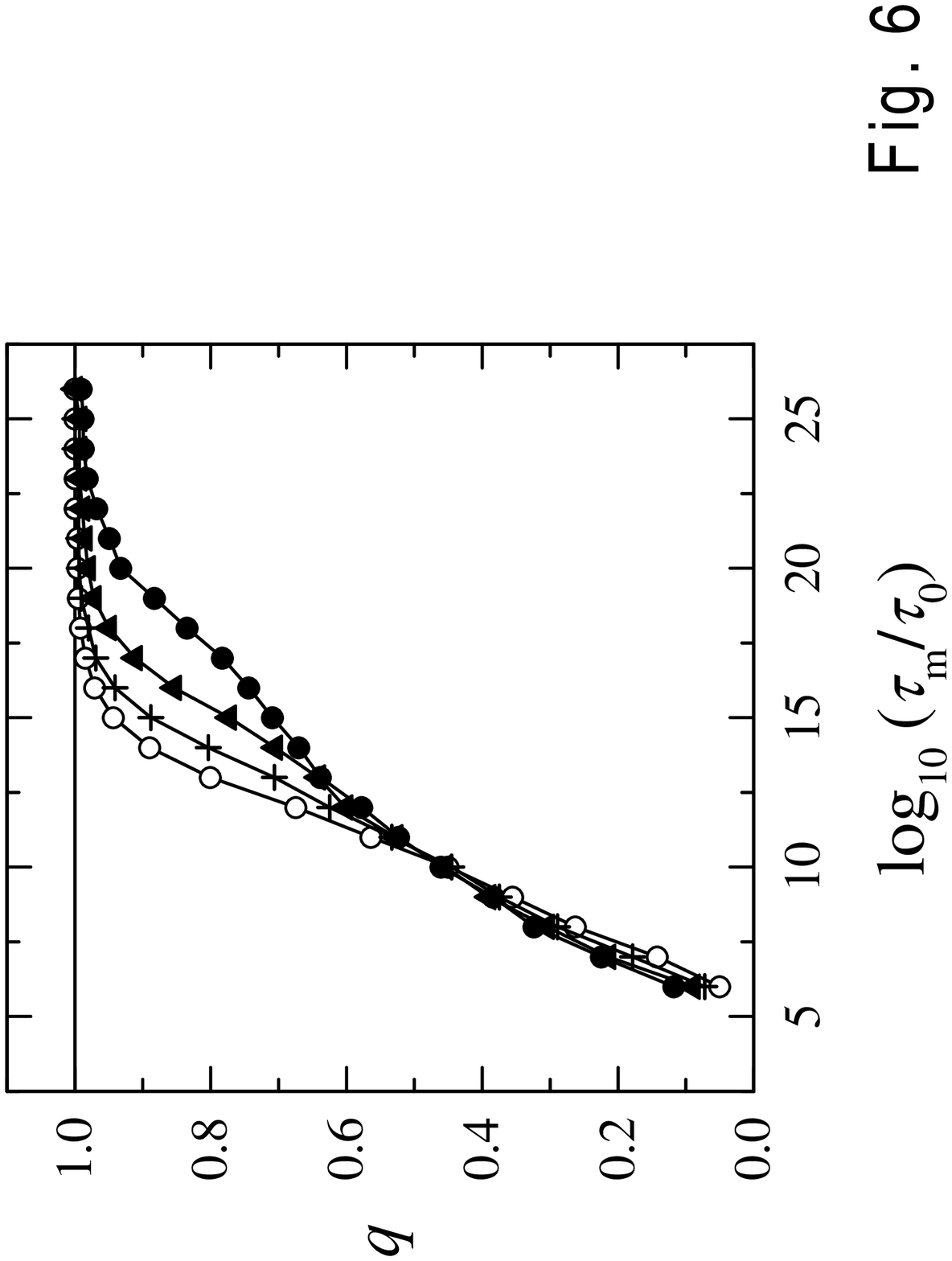}
\end{center}

\pagebreak
\epsfxsize=\hsize
\begin{center}
\leavevmode
\epsfbox{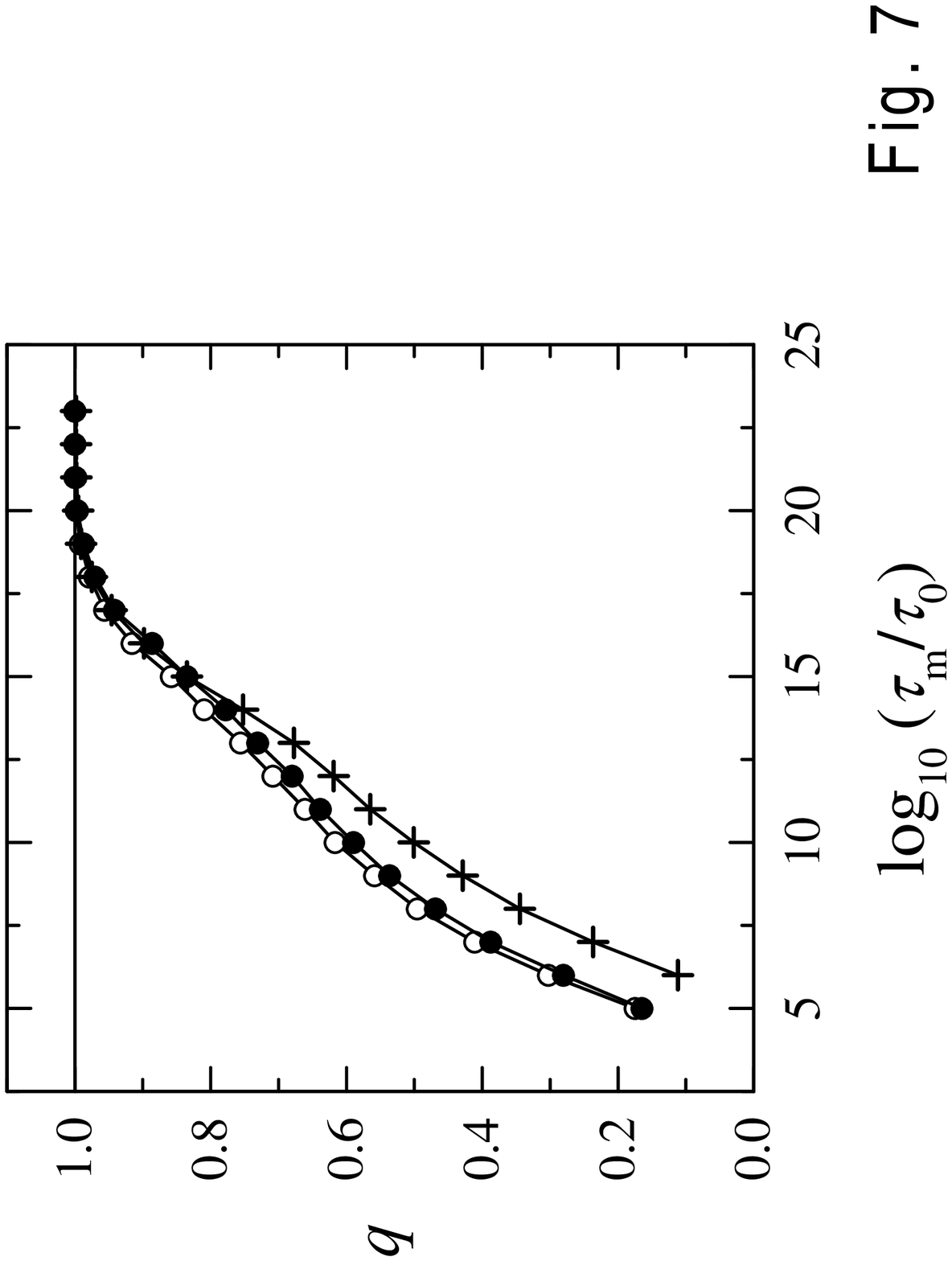}
\end{center}

\pagebreak
\epsfxsize=\hsize
\begin{center}
\leavevmode
\epsfbox{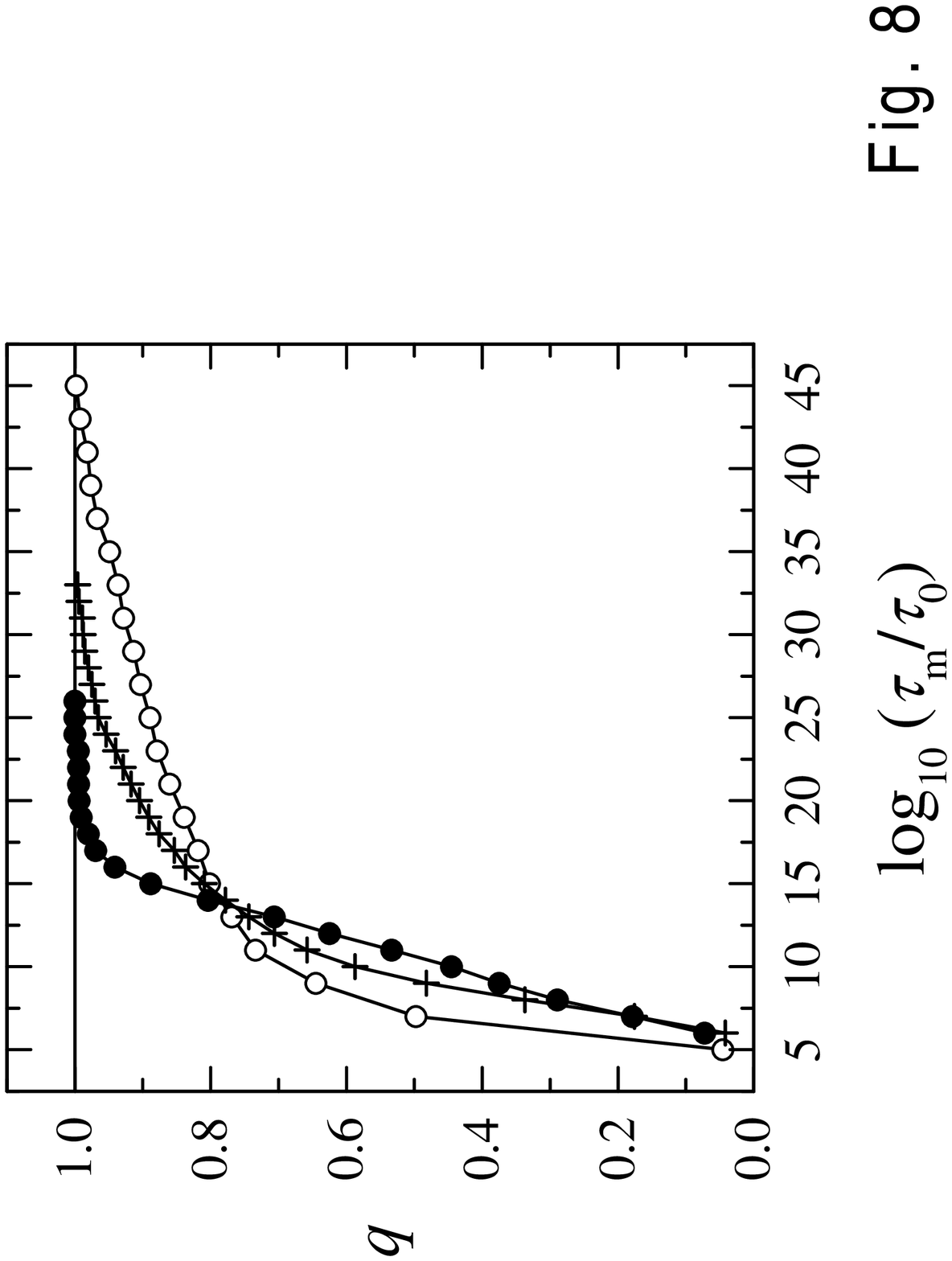}
\end{center}

\pagebreak
\epsfxsize=\hsize
\begin{center}
\leavevmode
\epsfbox{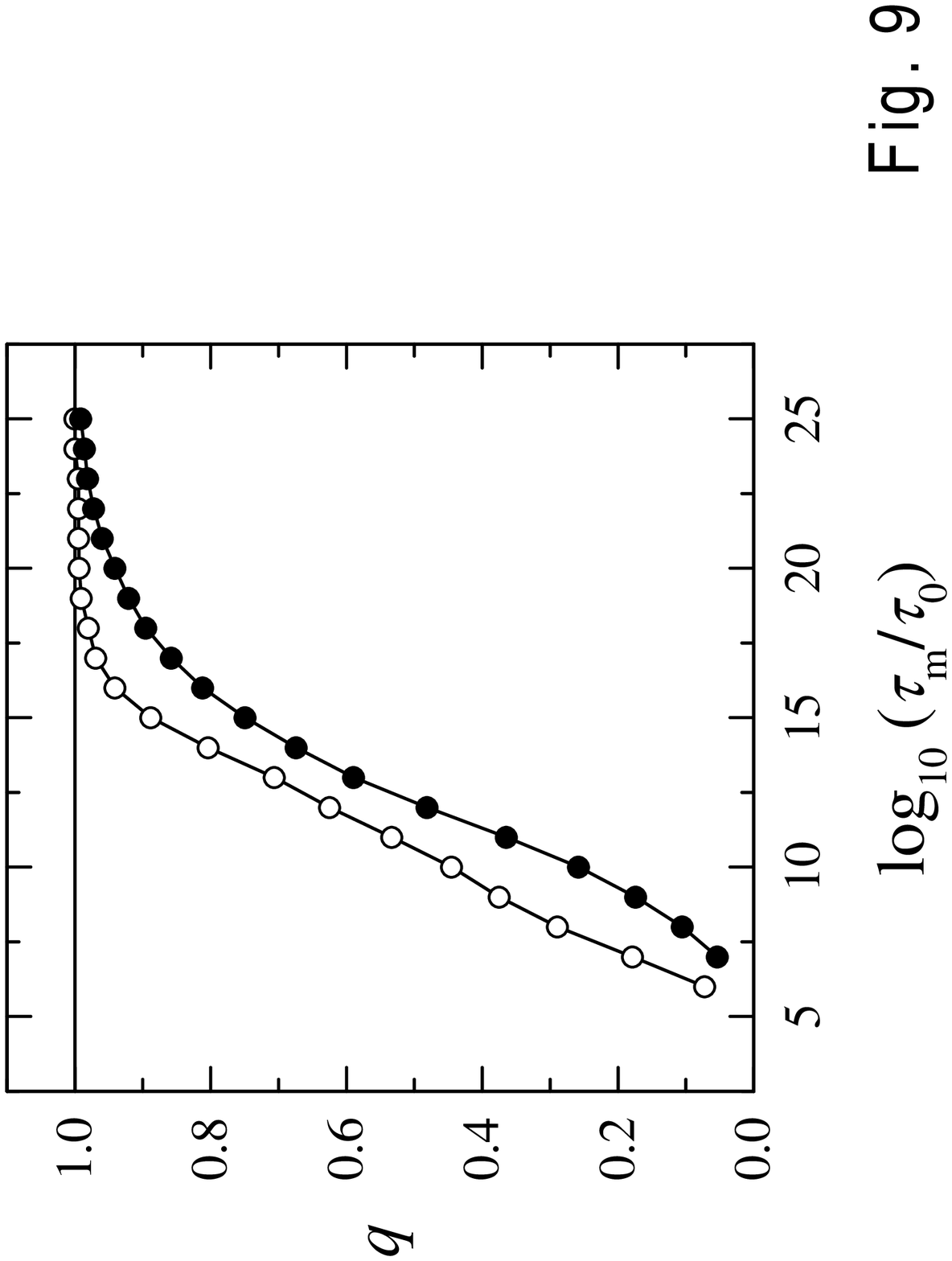}
\end{center}

\end{document}